# Stable Band-Gaps in Phononic Crystals by Harnessing Hyperelastic Transformation Media

Yan Liu [b], Zheng Chang [a, *], Xi-Qiao Feng [b]

[a] *College of Science, China Agricultural University, Beijing 100083, China*

[b] *AML & CNMM, Department of Engineering Mechanics, Tsinghua University, Beijing 100084, China*

**ABSTRACT**

The band structure in phononic crystals (PCs) is usually affected by the deformations of their soft components. In this work, hyperelastic transformation media is proposed to be integrated in the PCs' design, to achieve stable elastic band-gaps which are independent with finite mechanical deformations. For one-dimensional (1D) PCs, we demonstrate those with semi-linear soft component can keep all elastic wave bands unchanged with respect to external deformation field. While for those with neo-Hookean soft component, only S-wave bands can be precisely retained. An effective lumped mass method is presented to predict the transformation of the P-wave bands. Numerical simulations are performed to validate our theoretical predictions and the robustness of the proposed PCs. （113 words）

---

[*] Author to whom correspondence should be addressed. Electronic mail: changzh@cau.edu.cn

## 1. Introduction

Phononic crystals[1] (PCs) have attracted thorough attention, owing to their wide potential applications as filters, waveguides and sensors. In PC design, soft material always plays an important role for different purposes. For example, rubbery soft material is adopted in locally resonant sonic crystals[2] to enlarge the resonance effect. In recent days, the soft material becomes more enchanting for its high sensitivity to deformations[3], and also for its ability of undergoing reversible structural instability[4]. These essential features open a feasible route to achieving PC devices with tunable band-gaps[5].

However, on the other hand, stable band-gaps which are independent with the external stimuli are also pursued in some technologically important applications, in which the performance of the PCs is required to be robust enough to cope with harsh working environments, such as structural deformations and vibrations. Apparently, such requirement is a challenging issue for the PCs with soft components, since the above-mentioned features for soft materials become drawbacks in this circumstance.

Recently, hyperelastic transformation theory[6-8] has been proposed as a new route to manipulating elastic waves. More importantly, it reveals that hyperelastic transformation media, such as semi-linear materials[6] and neo-Hookean materials[7,8] can behave like smart transformation metamaterials[9] and process a space invariance for wave applications. Such properties shed light on the possibility to design PCs with soft components which have stable band-gaps under external mechanical stimuli.

In this Letter, by invoking hyperelastic transformation theory, we investigate the band-gap structures of PCs with hyperelastic transformation media as their soft component. Considering different types of hyperelastic transformation media, i.e. semi-linear materials and neo-Hookean materials, we propose a 1D PC which manifests unique or partial unique band structures for finite deformation of the soft components. We also consider a more realistic situation, in which the soft components of the PC subject to random mechanical deformation with various extent. Both theoretical analysis and numerical simulations demonstrate the robustness of the band structures for the proposed PCs.

## 2. One-dimensional phononic crystals with hyperelastic transformation media

Consider a 1D layered structure arranged alternately by a hard linear elastic material (A) and a soft hyperelastic material (B). In the initial configuration, as shown in Fig.1 (a), Part A and B intersect with each other in the same distance along the x-direction. The thicknesses of Part A and B are $a_A$ and $a_B$, respectively. As the soft component, Part B, is more sensitive to deformation than Part A, the external stimuli is considered to be applied only on Part B. For simplicity, the form of the stimuli is considered to be in the form of uniaxial tension in x-direction, as shown in Fig. 1(b).

To describe the linear elastic wave motion in hyperelastic materials, small-on-large theory[10] is utilized, in which the governing equation of elastic wave motion is

$$\nabla \cdot (\mathbf{C} : \nabla \mathbf{u}) = \rho \ddot{\mathbf{u}}, \quad (1)$$

in which $\mathbf{u}$ is the displacement vector, $\mathbf{C}$ and $\rho$ are the fourth-order tangent moduli tensor and the effective mass density, respectively, with the forms of

$$C_{ijkl} = J^{-1} F_{i\alpha} F_{k\beta} C_{0\alpha j\beta l}, \rho = J^{-1} \rho_0, \quad (2)$$

in which $F_{ij}$ is the component of the deformation gradient tensor $\mathbf{F}$ of the finite deformation, $J = \det(\mathbf{F})$, $\mathbf{C}_0$ and $\rho_0$ are the initial moduli and the initial mass density, respectively. $\mathbf{C}_0$ can be obtained from[10]

$$\mathbf{C}_0 = \frac{\partial^2 W}{\partial \mathbf{F} \partial \mathbf{F}}, \tag{3}$$

in which $W$ is the strain energy function of the soft media.

To achieve stable band-gaps, the strain energy function $W$ of Part B is considered to be in a semi-linear form[6]

$$W = \frac{\lambda_0}{2}(i_1 - 3)^2 + \mu_0((i_1 - 1)^2 - 2i_2 + 2), \tag{4}$$

in which $i_1$, $i_2$ are the invariants of deformation gradient $\mathbf{F}$, while $\lambda_0$ and $\mu_0$ are initial lame constants. According to hyperelastic transformation theory[6], the semi-linear strain energy function manifests an analogy between the pushing forward operation (Eq. (2)) in the small-on-large theory and the asymmetric transformation relations[11] in the traditional elastodynamic transformation theory[12], under the circumstances the deformation is rotation free. This means the wave responses of the deformed PC (Fig. 1 (b)) will be totally the same as the initial one (Fig. 1 (a)), implying the band structure of the PC is stable with respect to the applied deformations.

In a similar fashion, the strain energy function of part B can also be considered in a neo-Hookean form[8]

$$W = \frac{\lambda_0}{2}(J - 1)^2 - \mu_0 \ln J + \frac{\mu_0}{2}(I_1 - 3) \tag{5}$$

with $J = \det(\mathbf{F})$ and $I_1$ the first invariant of $\mathbf{U} = \sqrt{\mathbf{F}^T \cdot \mathbf{F}}$. However, as the analogy between the pushing forward relation and the asymmetric transformation relation stands only for S-waves[8], we cannot obtain a unique band structure by using neo-Hookean soft components. Nevertheless, in the band structure, all the S-wave bands can be expected to be identical between the initial and the deformed PCs. To analytically investigate the influence of deformation on the P-wave bands, an effective lumped mass method can be relied on, as shown in the Supplemental Material[13].

## 3. Numerical Simulations

To validate the theoretical prediction above, numerical simulations are performed by using the software COMSOL Multiphysics. A two-step model[8] is adopted to calculate the small-on-large wave motion. The first step is to calculate a static equilibrium equation for hyperelastic materials and further deduce the corresponding effective material parameter in deformed configuration. The second step is to calculate the band structure of the PC through a weak form PDE model.

To implement above scheme, a primitive cell is chosen in the deformed configuration, as shown in Fig.1 (c). The initial thicknesses of the Part A and B are set to be $a_A = a_B = 0.01\,\mathrm{m}$, while the height of the primitive cell is set to be $h = 0.02\,\mathrm{m}$. Periodic boundary conditions are imposed at the four boundaries surrounding the cell to calculate the band structure in the $\Gamma - X$ direction, as shown in Fig.1 (d). The left and right boundaries are set in Bloch periodicity conditions while continuity periodicity conditions are adopted in up and bottom boundaries.

The materials of part A and B are chosen to be aluminum ($\rho_A = 2730\,\mathrm{kg/m^3}$, $\lambda_A = 7.76 \times 10^{10}\,\mathrm{Pa}$ and $\mu_A = 2.87 \times 10^{10}\,\mathrm{Pa}$) and vulcanization rubber ($\rho_B = 1300\,\mathrm{kg/m^3}$, $\lambda_B = 1 \times 10^{6}\,\mathrm{Pa}$ and $\mu_B = 3.4 \times 10^{5}\,\mathrm{Pa}$). As mentioned earlier, for Part B, semi-linear and neo-Hookean strain energy function are considered. Meanwhile, Gent strain energy function with the form of

$$W = -\frac{\mu_0}{2} J_m \log(1 - \frac{I_1 - 3}{J_m}) - \mu_0 \log J$$
$$+ (\frac{\lambda_0}{2} - \frac{\mu}{J_m})(J - 1)^2 \quad (6)$$

is also considered as a comparison, in which $J_m$ is a dimensionless parameter related to the strain saturation of the material. Here we choose $J_m = 0.5$ in accordance with a previous literature[3].

The band structure for different soft components can be found in Fig.2. In the first numerical example, we calculate the band structure of the initial configuration (Fig. 1 (a)), as demonstrated in Fig.2 (a). The P-wave bands and S-wave bands are distinguished by red and blue lines. Three band-gaps can be observed in the range of [0 -2000] Hz, which are [732 -808] Hz, [914-1617] Hz, and [1691-1801] Hz, respectively.

For the deformed PCs, we consider the soft parts are uniaxial stretched with elongation strain $\varepsilon = 30\%$. As illustrated in Fig.2 (b), the band structure of the PC with deformed semi-linear component is exactly the same as that of the undeformed one (Fig.2 (a)), which is consistent with our theoretical prediction.

For the PC with neo-Hookean component, the S-bands (blue lines) are retained, just in consistency with our theoretical prediction. The P-bands (red lines) are changed due to the deformation, as shown in Fig.2 (c). In this sense, the second band-gap keeps unchanged with deformation, as both of the two boundaries of this band-gap are governed by S-bands. The first and third band-gaps change into the range of [701-808] Hz and [1691-1721] Hz, respectively.

As illustrated in Fig. 2(d), the PC with Gent component exhibits a totally different band structure. All the band-gaps shift to lower frequencies. Simultaneously, the bandwidths also manifest significant changes, in which the first band-gap is almost closed and the second and third band-gaps are changed by 42% and 56%, respectively.

In order to testify the correctness of the above band structures, the transmission spectra are also simulated for these cases. In the numerical processes, a super cell with eight primitive cells in x-direction is considered, as shown in Fig.1 (d). Instead of Bloch periodical boundary conditions, the left boundary of the super cell is set to be a harmonic wave source, while the right boundary is set to be an output terminal. The transmission coefficient **T** is defined by $T = 20 log_{10}(u_{out} / u_{in})$ in which $u_{in}$ and $u_{out}$ are the integrated displacement on the input and output terminals respectively. As the numerical results shown in Fig.2, the transmission spectra demonstrate a good agreement with the band structures for both the pass bands and band-gaps.

To systematically investigate the influence of deformation on the band structures of PCs with semi-linear or neo-Hookean components, the transformation of the band-gaps with the applied uniaxial tension is further investigated. For the PC with semi-linear component (not shown), all the bands keep unchanged no matter how large the extent of the uniaxial tension is. For the neo-Hookean case, as demonstrated in Fig. 3, all the S-bands are independent with the uniaxial tension. The first P-wave band shifts to a lower frequency, making the first band-gap increase its width for about 49% when $\varepsilon = 40\%$. Meanwhile, the third band-gap is observed to reduce its width with deformation nearly up to closure occurring at $\varepsilon = 40\%$. The effective lumped mass method[13] is also adopted to analyze the transformation of P-bands with the applied deformation. As shown in Fig. 3, the analytical results match well with the results from numerical simulations.

In the following, a more realistic situation is investigated by using transmission spectrum, in which the soft components of the PCs subject to randomly different mechanical deformations with $\varepsilon \in [0, 30\%]$. With the super cell mentioned above (Fig. 1(d)), randomly different uniaxial tension is applied on each area of the soft component. Three different randomly generated cases are considered, as shown in Fig. 4(a). For the PCs with semi-linear component, the transmission spectra for the three cases (not shown) totally coincides with each other and is all the same as that of the undeformed PC (not shown). For the PCs with neo-Hookean components, as illustrated in Fig. 4(b), all the boundaries governed by S-bands, which are the right boundary of the first band-gap, the both boundaries of the second band-gap, together with the left boundary of the third band-gap, agree very well with each other. Although not identical, the remaining two band-gap boundaries distribute in a very close frequency range. The shaded areas in Fig. 4(b) demonstrates the band-gaps calculated by averaging the global deformation of the super cell. The good agreement among the transmission spectra (lines) and the band-gap areas (shaded area) implying that the band-gaps for PCs with neo-Hookean components can be well predicted by considering the overall deformation states of the PCs. Moreover, the results also illustrated that stable band-gaps can also be expected in such a realistic circumstance.

## 4. Discussions and conclusions

Apparently, from above theoretical and numerical analysis, semi-linear material exhibits significant advantages in designing PCs with fully stable band-gaps. However, till now, the semi-linear soft material can hardly be found in nature. Moreover, the feature of spatial equivalence that semi-liner materials demonstrated strictly requires the deformation is rotation free[6]. These issues make it very tough to practically realize PCs with stable band-gaps, by using semi-liner materials. On the other hand, although can only result in stable S-wave bands, the neo-Hookean material are very common in nature, and its feature of spatial equivalence is independent with any deformation states[8]. In our analysis, as shown in Fig. 3, although the third band-gap may be closed with a large deformation applied on the soft part, the width of first band-gap even increased with the deformation. The investigation of randomly different deformations in the PCs also demonstrates the robustness of the performance of the neo-Hookean material. Therefore, we propose that the neo-Hookean material may be currently the most potential candidate to realize PC devices with stable band structures.

In conclusion, in this work, we integrated hyperelastic transformation theory in the PC design to achieve stable band structures which are independent with the external deformation. With a 1D PC with pre-deformed semi-linear component, we theoretically and numerically demonstrated the band structure are identical with that of the initial PC configuration. The band structures of PCs with pre-deformed neo-Hookean component are also systematically investigated, showing that the deformation has no effects on S-bands, but has considerable effects on P-bands. The analysis on the relation between the transformation of band-gaps and the deformations applied implies that the neo-Hookean material can be utilized to implement PCs with stable band-gaps under certain reasonable circumstances. This work can be seen as another application of hyperelastic transformation theories, as well as the elastodynamic cloak[6, 7], elastic wave mode splitter[8], and shear wave beam band[14]. More importantly, the proposed PCs may have significant applications in the fields where high precision transmission and measurement are required.

**Acknowledgments**

This work was supported by the National Natural Science Foundation of China (grant number 11602294).

**Figures**

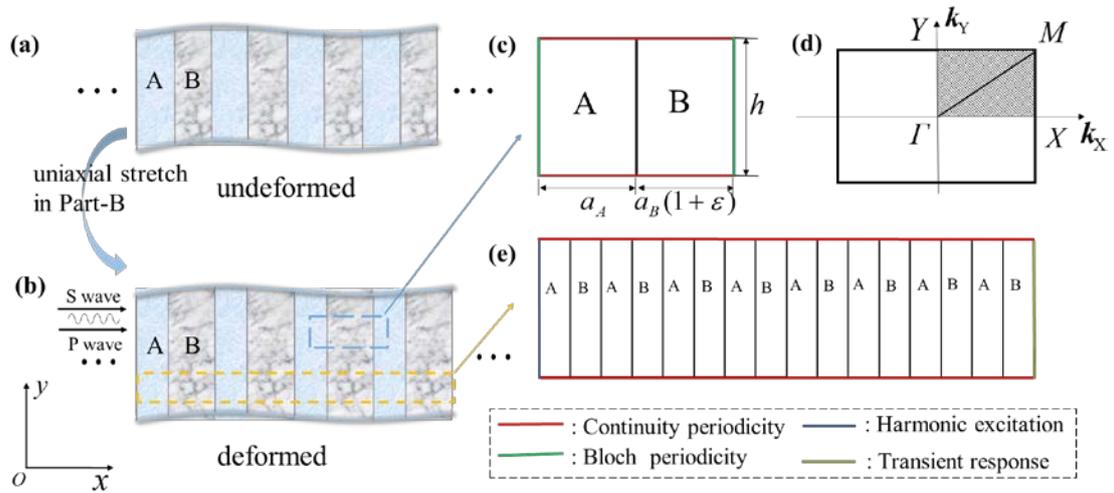

**Fig.1** Schematics of the 1D PC model. (a) The initial configuration of the 1D PC. (b) The deformed configuration. (c) A primitive cell of the PC to predict the elastic band structures. (d) First Brillouin zone for the primitive cell. In this work, only the band structure on Γ-X is considered. (e) A super cell consisting of eight primitive cell applied to calculate the transmission spectrum of the PC.

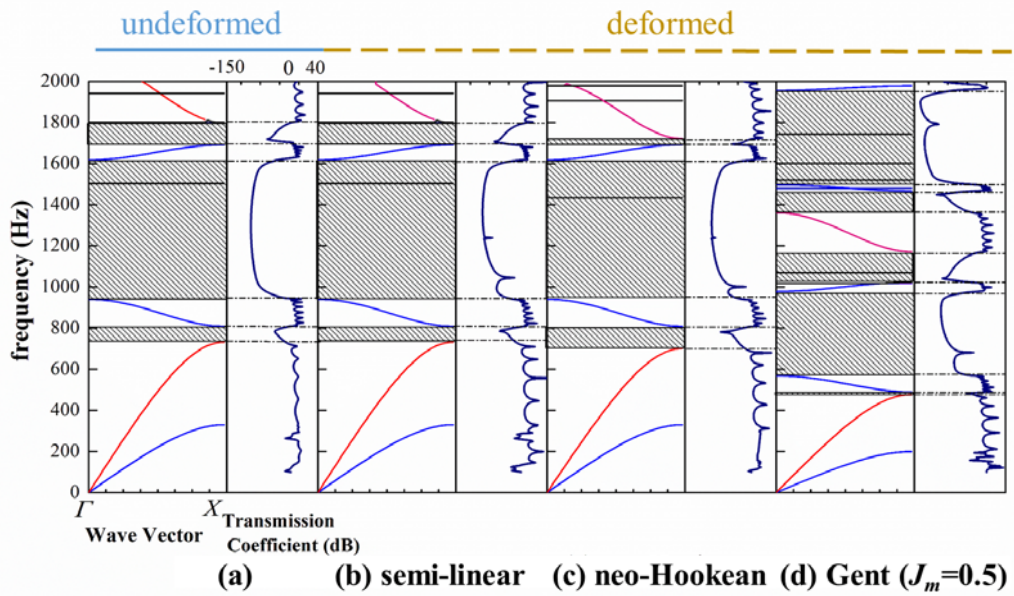

**Fig.2** Band structures and transmission spectrums for the PCs with different soft components. (a) Undeformed soft component, which is also equivalent to a linear elastic material with initial material parameters. (b)-(d) Semi-linear, neo-Hookean, and Gent ($J_m=0.5$) soft components, respectively. In (b)-(d), all the soft components are applied a uniaxial stretched with elongation strain $\lambda=0.3$. In the band structures, the P-bands and the S-bands are distinguished by red and blue lines. The band-gaps are denoted by the shaded areas.

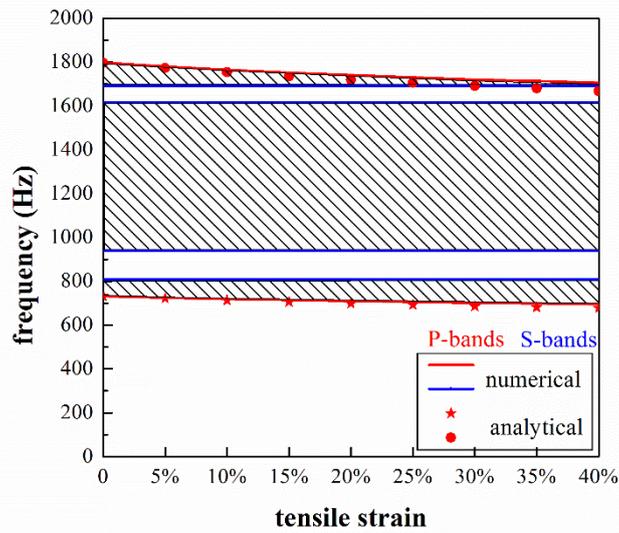

**Fig.3** Elastic band-gaps vs the tensile strain applied on the neo-Hookean component of the PC. The band-gaps are denoted by the shaded areas. The band-gap boundaries governed by P-bands and the S-bands are distinguished by red and blue colors, respectively. The lines represent the numerical simulation results while the scatters are the analytical solutions.

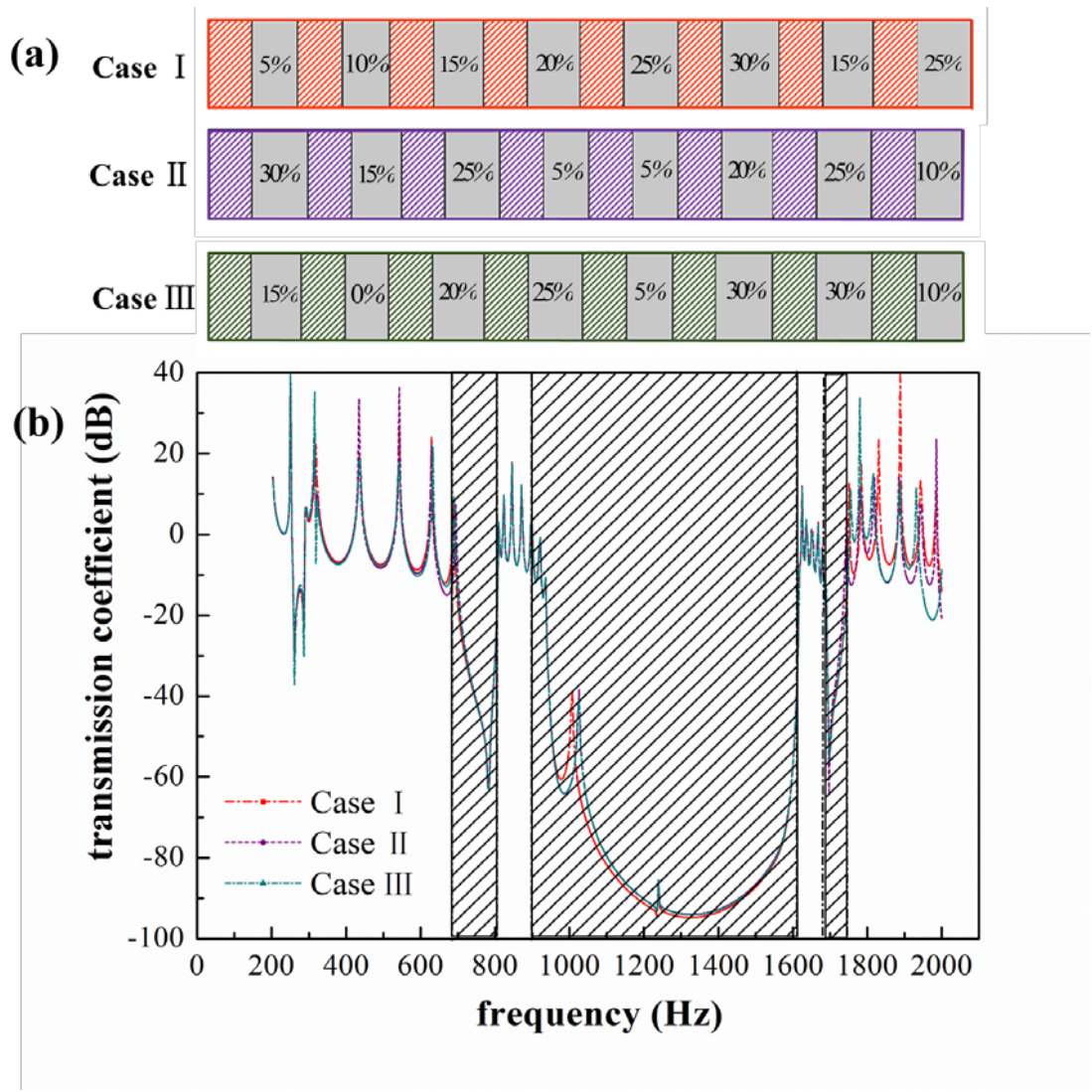

**Fig.4** Transmission spectra for the PC with neo-Hookean soft component subject to randomly different mechanical deformations. (a) Three randomly generated cases represent the elongation strains applied on different areas of the soft component. (b) The transmission spectra for the three cases illustrated in (a). The shaded areas denotes the band-gaps predicted by the averaging the global deformation of the super cell.